\documentclass{aa}        
\usepackage{graphics}
\usepackage{epsfig}
\begin{document}
\thesaurus{10.19.3, 13.09.3, 09.03.1, 09.08.1, 09.13.2, 13.9.3}

\title
{The radial distribution of OB star formation in the Galaxy  
\thanks{Based partly on results collected at the European Southern  
Observatory, La Silla, Chile}}

\author{L.~Bronfman\inst{1} \and S.~Casassus\inst{1,2} 
\and J.~May\inst{1} \and L.--\AA.~Nyman\inst{3,4}}
\offprints{L.~Bronfman}
\institute{Departamento de Astronom\'{\i}a, Universidad de Chile, 
Casilla 36-D, Santiago, Chile \and Department of Physics, University of
Oxford, Keble Road, Oxford OX1 3RH, England \and SEST, ESO-La Silla, Casilla
19001, Santiago 19, Chile \and Onsala Space Observatory, S--439 92 Sweden}
\date{received;      accepted} \authorrunning{Bronfman et al.}
\titlerunning{Radial distribution of OB star formation in the Galaxy} \maketitle
\begin{abstract}

We present the azimuthally averaged radial distribution of 748 regions of OB star formation in the whole galactic disk, based on our previous CS(2-1) survey of UC H II regions. Embedded massive stars produce a total FIR luminosity of $1.39\, 10^{8}\, L_{\odot}$ within the range $0.2  \leq R/R_{\circ} \leq 2 $ in galactocentric radius. We find 492 massive star forming regions within the solar circle, producing $81\%$ of the total FIR luminosity. Separate analyses of the 349 sources in the I and II quadrant (north), and of the 399 sources in the III and IV quadrant (south), yield FIR luminosities (extrapolated to the complete galactic disk) of $1.17\, 10^{8}\, L_{\odot}$ and of $1.60\, 10^{8}\, L_{\odot}$, respectively. Massive star formation is distributed in a layer with its centroid $Z_{\circ}(R)$ following that of molecular gas for all galactocentric radii, both north and south. Its thickness for $R\,\leq\,R_{\circ}$ is $\sim 73$ pc (FWHM), $62\%$ the thickness of the molecular gas disk. The FIR luminosity produced by massive stars has a well defined maximum at $R = 0.55 \,R_{\circ}$, with a gaussian FWHM of $0.28\,R_{\circ}$ - compared with $0.51\,R_{\circ}$ for the H$_2$ surface density distribution. Toward the outer Galaxy, down from the maximum, the face-on FIR surface luminosity decays exponentially with a scale length of $0.21 \,R_{\circ}$, compared with $0.34 \,R_{\circ}$ for the H$_2$ surface density.  Massive star formation per unit H$_2$ mass is maximum for $R\,\sim0.55\,R_{\circ}$ in the southern Galaxy, with a FIR surface luminosity to H$_2$ surface density ratio of  $\sim 0.41\, L_{\odot}/M_{\odot}$, compared with $\sim 0.21\, L_{\odot}/M_{\odot}$ at the same radius in the north, and with an average of $\sim 0.18\, L_{\odot}/M_{\odot}$ for the whole galactic disk within the solar circle.   

\keywords{Galaxy:
structure --- infrared: interstellar: continuum --- interstellar medium :
clouds --- interstellar medium: H II regions  --- interstellar medium:
molecules, millimeter lines --- radio lines:  molecular} \end{abstract}

\section{Introduction}

%figure1
\begin{figure*}
\begin{center}
\resizebox{\textwidth}{!}{\epsfig{file=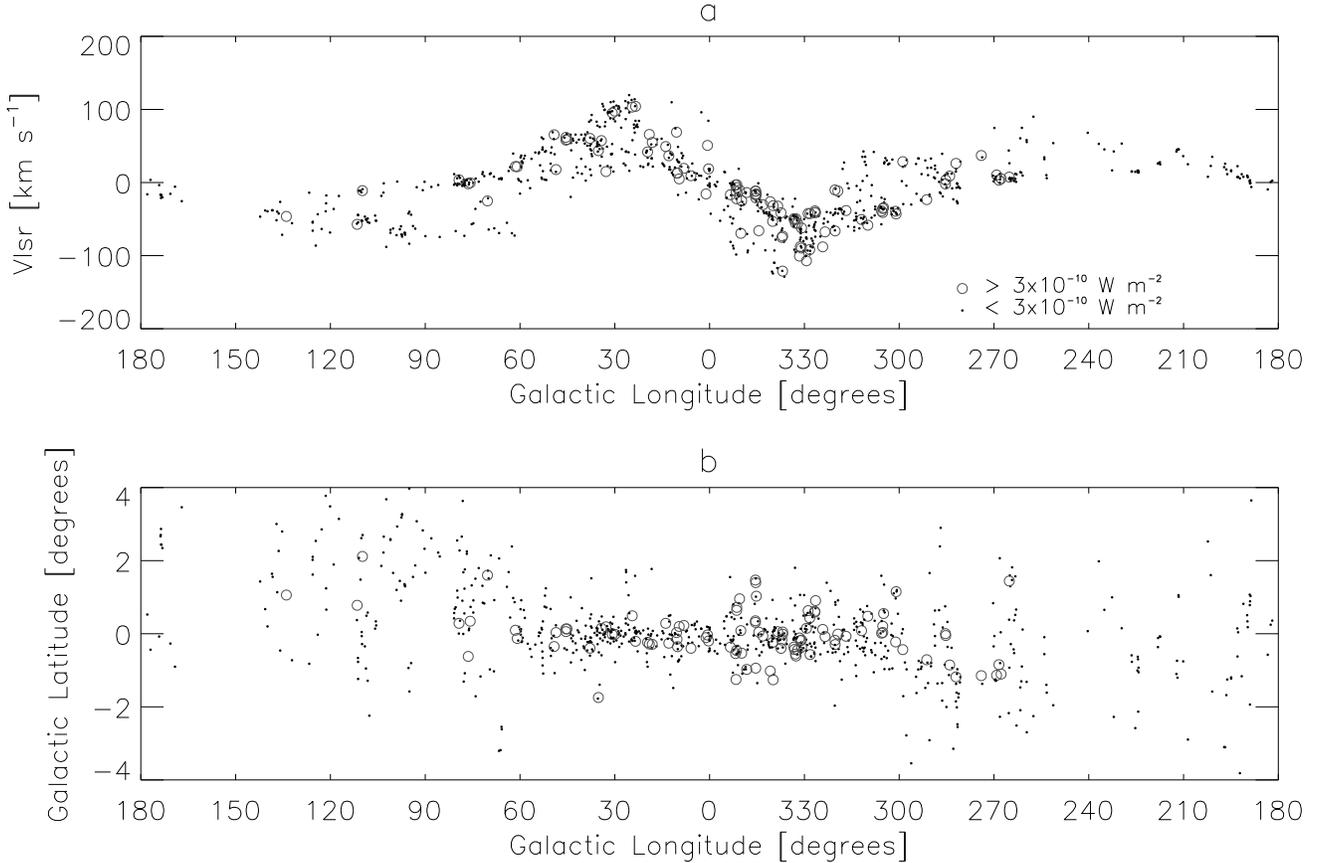}}
\end{center}
\caption{\small Distribution of massive star forming regions in the whole Galaxy: 
(a) Longitude-velocity; and (b) Longitude-latitude. Sources with FIR
fluxes higher than $3 \, 10^{-10}\, {\rm W}\,{\rm m}^{-2}$, corresponding to
an O8 star with a luminosity of $1.7\, 10^{5}\,L_{\odot} $ at a distance of 4.25 kpc, are represented with open circles}  
\end{figure*}

How many regions of massive star formation are there in the galactic disk and how 
are they distributed? Do all these regions have similar luminosities? Dust
extinction prevents optical observations of OB associations farther than 3 kpc
from the Sun. An alternative way to study the population of recently formed OB
stars in the Galaxy is to use  observations at longer wavelenghts in the far
infrared (FIR), where the ISM is more transparent. Ultraviolet light from OB
stars, born in the dense dusty cores of giant molecular clouds, ionizes the
closeby gas creating ultracompact (UC) H II regions and heating the
surrounding dust, which reradiates the energy mostly in the FIR. The
star-forming region becomes thus detectable by IRAS, generally as a point-like
source. But which of the many IRAS point-like sources in the galactic plane are
heated by embedded massive stars? The FIR spectral characteristics of UC H II
regions identified with the VLA have been studied by Wood \& Churchwell
(1989a), who calibrated their FIR colours as a tool for their identification.
Applying their criterion to the IRAS point-like sources in the galactic plane,
they found 1646 sources which, according to their IRAS 60$\mu$/12$\mu$ and
25$\mu$/12$\mu$ flux ratios, are candidates to be sites of OB star formation
containing UC H II regions (Wood \& Churchwell 1989b; WC89).    

Millimeter-wave emission lines from dense gas cores, where OB stars form, 
can be  used to obtain their kinematic distances and derive the FIR luminosity of their associated dust, heated by the embedded stellar sources. The $\rm
^{12}CO(1-0)$ line has been surveyed toward IRAS point-like sources in the
outer Galaxy (Wouterloot \& Brand 1996; and references therein), using less
restrictive FIR colors criteria. Within the solar circle, however, where the
majority of massive star formation takes place, there are several molecular
clouds at different distances for most lines of sight, yielding complex CO
velocity profiles. The $\rm ^{12}CO(1-0)$ line, affected by large optical
depths, will tell us little about the dense interior of molecular clouds. The
$\rm ^{13}CO(1-0)$ and $\rm C^{18}O(1-0)$ lines, of intermediate and low
opacity in the galactic plane, are still sensitive to molecular gas column
density, which can be enhanced not only toward  dense cores but also toward
extended regions of intermediate density.     

One of the best specific tracers of dense gas is the CS molecule. Its
$J=(2-1)$ rotational line, at a frequency of 99 GHz, has an excitation density
threshold of $\rm 10^{4}$ - $10^{5}$ cm$^{-3}$ and is ubiquitous toward regions
of massive star formation but seldom elsewhere in the galactic disk - with the
exception of the galactic center region (Bally et al. 1987). A complete survey
of the $\rm CS(2-1)$ line toward IRAS point-like sources with FIR colors of UC
H II regions in the galactic plane (Fig. 1) was presented by Bronfman et al.
(1996; BNM). Nearly in all cases the $\rm CS(2-1)$ profiles obtained in such
survey are singular; the dense gas cores are associated one-to-one with the
heated dust which produces the FIR emission detected by IRAS as a point-like
source. The $\rm CS(2-1)$ line, as well as other higher frequency rotational
lines of CS, have been used successfully to map dense molecular gas cores and
to derive their physical parameters (Plume et al. 1997; and references therein).

%figure2
\begin{figure*}
\begin{center}
\resizebox{\textwidth}{!}{\epsfig{file=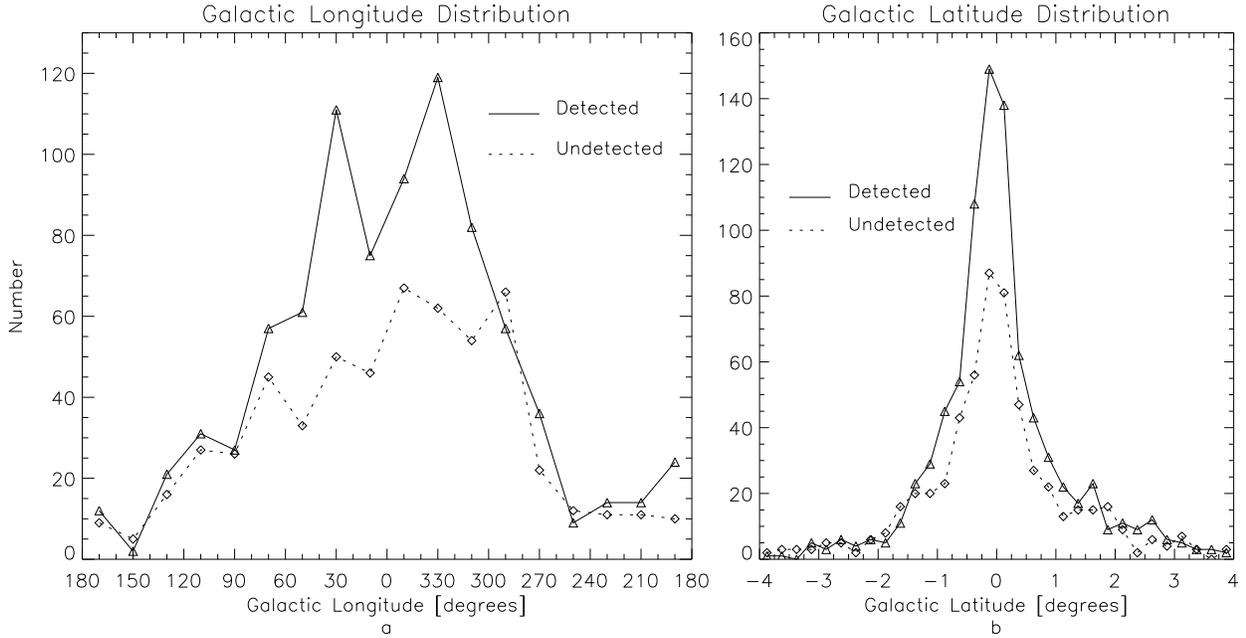}}
\end{center}
\caption{\small Distribution of detected and undetected sources integrated in 
(a) galactic latitude and (b) galactic longitude} 
\end{figure*}

The main goal of the present paper is to derive the mean radial distribution of the face-on FIR surface luminosity produced by embedded massive stars in the whole Galaxy, and to compare it with the mean radial distribution of H$_2$ surface density as derived from CO surveys of the Milky Way. For a direct comparison, thus, we analyze our sample of OB star formation regions in a manner as close as possible as that used for the derivation of the H$_2$ surface density (Bronfman et al. 1988; Paper I). 

As an initial step in the analysis of the BNM CS$(2-1)$ survey, we present 
the azimuthally averaged distribution of regions of massive star formation along the whole galactic disk, $0.2 \leq R/R_{\circ} \leq 2$. We also perform separate analyses of the northern (I and II quadrants) and southern (III and IV
quadrants) Galaxy in order to search for large scale deviations from axial
symmetry. Using the kinematic distances derived from the CS$(2-1)$ line
profiles and the FIR fluxes from the  IRAS Point Source Catalog (PSC), we evaluate the FIR face-on surface luminosity, averaged over galactocentric rings $0.1\, R_{\circ}$ (0.85 kpc) wide, for the northern, southern, and complete galactic disk. The results obtained are compared with the molecular hydrogen surface density distribution as derived from large scale $\rm CO(1-0)$ surveys of the Galaxy. 

In  Sect. 2 we summarize the observations and the results of 
the BNM $\rm CS(2-1)$ survey. In Sect. 3 we describe our axisymmetric analysis and derive the mean radial distribution of OB star formation in the Galaxy, the azimuthally averaged face-on FIR surface luminosity these stars produce, and 
the mean FIR luminosity per OB star forming region. We evaluate the shortcomings in our analysis which are due to the spiral structure of the Galaxy and to the incompleteness of our sample. In Sect. 4 we compare these results with the surface density of cold molecular gas as traced by CO, discuss the large scale asymmetries of OB star formation in the Galaxy, and estimate the
contribution of embedded OB stars to the FIR luminosity of Giant Molecular Clouds (GMCs). In Sect. 5 we summarize our conclusions.

\section{Observational results}

The BNM $\rm CS(2-1)$ survey  was carried out between 1989 and 1992 with the 
15-m SEST (Swedish-ESO Submillimeter Telescope) at La Silla, Chile, and with
the OSO (Onsala Space Observatory) 20-m Telescope, Sweden, in May 1993.The
angular resolution of the telescopes at 99 GHz, the CS(2-1) line frequency, is
$50 \arcsec$ for SEST and $39 \arcsec$ for OSO (main beam FWHM), similar to the
IRAS survey position accuracy. The instruments, observing techniques, and data
calibration are described by BNM. The observed sources were selected from the
IRAS PSC (Version 1), using exactly the same criterion as
in WC89.  They are all located near the galactic plane,
they have a $\rm 25\mu/12\mu$ flux ratio larger than 3.7 and a $\rm 60\mu/25\mu$
flux ratio larger than 19.3, and are not identified as stars or galaxies. A
total of 1427 sources were selected. The survey covered $ -2^{\circ}\leq b\leq
+2^{\circ}$ for longitudes corresponding to the inner Galaxy, $ 300^{\circ}\leq
l\leq 360^{\circ}$ and $ 0^{\circ}\leq l \leq 60^{\circ}$. For the rest of the
galactic plane the latitude coverage was $-4^{\circ}\leq b \leq +4^{\circ}$.  

Of the 1427 sources observed, the survey detected 843 (59\%) above a level of 
$\sim\, 0.3$ K peak T$_{\rm A}$ (BNM), with a few detections below this
level, particularly in the far outer Galaxy (Table 1). Of the total number of
sources detected, 37\% are in the I quadrant; 10\% in the II quadrant; 11\% in
the III quadrant, and 42\% in the IV quadrant. The fraction of detected sources
is 0.58 toward the outer Galaxy (II and III quadrants), and 0.60 toward the
inner Galaxy (I and IV quadrants); toward certain longitudes in the direction
of the molecular annulus, however, the number of detected sources more than
doubles that of undetected ones. 

The distribution in longitude of the detected sources, integrated over latitude, shows two prominent maxima around $l = 30^{\circ}$ and $l =  330^{\circ}$,
toward lines of sight tangent to a galactocentric circle $0.5\, R_{\circ}$ in
radius (Fig. 2a). These maxima, symmetric with respect to $l = 0^{\circ}$, are
strong direct observational evidence  for the existence of an annulus of
massive star formation in the Milky Way that is coincident with the molecular
annulus, as traced by CO. The center of the distribution in galactic latitude 
of the detected sources (Fig. 2b) is very close to $b = 0\degr$, with an average of  $ -0{{\rlap.}{^{\circ}}}05\pm 0 {{\rlap.}{^{\circ}}}03$ and a FWHM of 
$ \sim 0 {{\rlap.}{^{\circ}}}8 $. The IRAS point-like sources not detected in CS, in
comparison, are more evenly distributed in the galactic disk, and their
latitude distribution is about 20\% wider. 

\section{Mean radial distribution of embedded OB stars}

%table1
\begin{table}
\
\caption{Distribution in galactocentric radius of the sources included
in the analysis}
\
\begin{tabular}{l@{\hspace{1.25cm}}l@{\hspace{1cm}}l@{\hspace{1cm}}l}
        \hline
        \noalign{\smallskip}
        $R_{\rm bin}$ & $N_{\rm North^{\rm a}}$ & $N_{\rm South^{\rm a}}$ &
Sampled area\\ 
        $(R_{\rm o})$ & & & (kpc$^{2}$)\\ 
        \noalign{\smallskip}
        \hline
        \noalign{\smallskip}
      0.25 & 1\,\,\,\,\,\, (1) & 0 \,\,\,\,\,\,(0) & 2.88\\
      0.35 & 2\,\,\,\,\,\, (2) & 6 \,\,\,\,\,\,(6) & 5.32\\
      0.45 & 13\,\,\, (13) & 9 \,\,\,\,\,\,(9) & 7.64\\
      0.55 & 48\,\,\, (48) & 42\,\,\, (41) & 9.93\\
      0.65 & 40\,\,\, (38) & 64\,\,\, (60) & 12.21\\
      0.75 & 55\,\,\, (52) & 61\,\,\, (57) & 14.49\\
      0.85 & 34\,\,\, (25) & 39\,\,\, (31) & 16.77\\
      0.95 & 44\,\,\, (15) & 34\,\,\, (18) & 19.04\\ 
      1.05 & 32\,\,\, (18) & 59\,\,\, (24) & 21.22\\
      1.15 & 9\,\,\,\,\,\, (4) & 37\,\,\, (24) & 23.30\\
      1.25 & 11\,\,\, (6) & 16\,\,\, (9) & 25.38\\
      1.35 & 25\,\,\, (22) & 14\,\,\, (8) & 27.46\\
      1.45 & 18\,\,\, (16) & 6 \,\,\,\,\,\,(3) & 29.54\\
      1.55 & 12\,\,\, (7) & 4 \,\,\,\,\,\,(4) & 31.62\\
      1.65 & 2\,\,\,\,\,\, (2) & 5 \,\,\,\,\,\,(4) & 33.70\\
      1.75 & 1\,\,\,\,\,\, (1) & 1 \,\,\,\,\,\,(4) & 35.78\\
      1.85 & 1\,\,\,\,\,\, (1) & 2 \,\,\,\,\,\,(2) & 37.87\\
      1.95 & 1\,\,\,\,\,\, (1) & 0 \,\,\,\,\,\,(0) & 39.95\\
      Total & 349 (272) & 399 (304) & 394.1\\
      \noalign{\smallskip}
    \hline
\end{tabular}

\begin{list}{}{}
\item[$^{\rm a}$] In parenthesis the number of sources when a lower luminosity 
cutoff is applied to the sample (see Sect. 3.5)   
\end{list}
\end{table}

%table2
\begin{table*}
\
\caption{The embedded massive stars layer}
\[
\begin{array}{l@{\hspace{0.5cm}}l@{\hspace{0.5cm}}l@{\hspace{0.5cm}}l@
{\hspace{0.5cm}}l@{\hspace{0.5cm}}l@{\hspace{0.5cm}}l@{\hspace{0.5cm}}
l@{\hspace{0.5cm}}l@{\hspace{0.5cm}}l}
\hline
\noalign{\smallskip}
R_{\rm bin} & N_{\rm North} & N_{\rm South} & N_{\rm Comb} & Z_{\rm o\, 
North} & Z_{\rm o\, South} & Z_{\rm o\, Comb} & Z_{\frac{1}{2}\, \rm North} &
Z_{\frac{1}{2}\, \rm South} & Z_{\frac{1}{2}\, \rm Comb}\\ (R_{\rm o}) & ({\rm
kpc}^{-2}) & ({\rm kpc}^{-2}) & ({\rm kpc}^{-2}) & ({\rm pc}) & ({\rm pc}) &
({\rm pc}) & ({\rm pc}) & ({\rm pc}) & ({\rm pc})\\ \noalign{\smallskip} \hline
\noalign{\smallskip} 0.25	& 	0.35	\pm	0.35	&    0.35 \pm 
0.35	&	0.17	\pm	0.35	&	&	&	&	&
&	\\	 0.35	&	0.38	\pm	0.27	&	1.13	\pm
0.46	&	0.75	\pm	0.53	&              	&	-11  \pm
1	&	-8	\pm	1	&	&	33 	\pm  1  &  28  
\pm 1   \\  0.45	&	1.70	\pm	0.47	&	1.18	\pm
0.39	&	1.44	\pm	0.61	&	+5	\pm	1	&
+5	\pm	1	&	+5	\pm	1	&	30	\pm
1	&	12	\pm	1	&	24	\pm	1\\ 0.55
&	4.83	\pm	0.70	&	4.23	\pm	0.65	&	4.53
\pm	0.96	&	-6	\pm	1	&	+6	\pm	1
&	-1	\pm	1	&	30	\pm	2	&	 46 \pm
  3	&	38	\pm	2\\ 0.65	&	3.28	\pm	0.52
&	5.24	\pm	0.66	&	4.26	\pm	0.83	&	-1
\pm	1	&	-18 	\pm	1	&	-12 	\pm	1
&	49	\pm	3 &	     32\pm          	3	&	42
\pm	3\\ 0.75	&	3.80	\pm	0.51	&	4.21	\pm
0.54	&	4.00	\pm	0.74	&	-19  \pm	3	&
+2	\pm	1	&	-8	\pm	1	&	41	\pm
10	&	48	\pm	3	&	46	  \pm 5\\ 0.85	&
2.03	\pm	0.35	&	2.33	\pm	0.37	&	2.18	\pm
0.51	&	+6	\pm	4	&	+13  \pm	7	&
+11	\pm	4	&	42	\pm	11	&	 74    \pm  41 
 &    62   \pm  21\\ 0.95	&    2.31 \pm	0.35	&	1.79	\pm
0.31	&	2.00	\pm	0.46	&	+9	\pm	4	&
-22  \pm	7	&	-4	\pm	4	&	30	\pm
10	&	63	  \pm  18  &  51  \pm  11\\ 1.05	&	1.51
\pm	0.27	&	2.78	\pm	0.36	&	2.31	\pm	0.47
&	+50 	\pm	7	&	-24 	\pm	3	&	+2
\pm	3	&	75	\pm     18   & 70 \pm 7 & 83 \pm 11\\ 1.15
&	0.39	\pm	0.13	&	1.59	\pm	0.26	&	0.99
\pm	0.29	&	+77 	\pm	5	&	-83 	\pm	4
&	-52 	\pm	4	&	154	\pm	6	&	86
\pm	5	&	128	\pm	4\\ 1.25	&	0.43	\pm
0.13	&	0.63	\pm	0.16	&	0.53	\pm	0.20	&
+9	\pm	5	&	-84 	\pm	4	&	-46 	\pm
3	&	141	\pm	13	&	94	\pm	9	&
127	\pm	9\\ 1.35	&	0.91	\pm	0.18	&	0.51
\pm	0.14	&	0.71	\pm	0.23	&	+48 	\pm	3
&	-103	\pm	7	&	-7	\pm	3	&	132
\pm		6&	133	\pm	8	&	157	\pm	5\\
1.45	&	0.61	\pm	0.14	&	0.20	\pm	0.08	&
0.41	\pm	0.17	&	+243	\pm	6	&	-47 	\pm
18	&	+171	\pm	7	&	121	\pm	13	&
142	\pm	27	&	195	\pm	21\\ 1.55	&	0.38
\pm	0.11	&	0.13	\pm	0.06	&	0.25	\pm	0.13
&	+298	\pm	7	&	-60 	\pm	5	&	+208
\pm	6	&	218	\pm	7	&	71	\pm	5
&	265	\pm	6\\ 1.65	&	0.06	\pm	0.04	&
0.15	\pm	0.07	&	0.10	\pm	0.08	&     	  	   
&	-60 	\pm	34	&	+148	\pm	25	&      	&
189	\pm	82	&	420	\pm	48\\ 1.75	&	0.03
\pm	0.03	&	0.03	\pm	0.03	&	0.03	\pm	0.04
&	&	&	&	&	&      \\ 1.85	&	0.03	\pm
0.03	&	0.05	\pm	0.04	&	0.04	\pm	0.05	&
&	&	&	&	&      \\ 1.95	&	0.03	\pm	0.03
&	0.05 \pm  0.04	&	0.01	\pm	0.03	&	&	&
&	&	&	\\ 
\noalign{\smallskip} 
\hline 
\end{array} 
\]
\end{table*}

%figure3
\begin{figure*}
\begin{center}
\resizebox{\textwidth}{!}{\epsfig{file=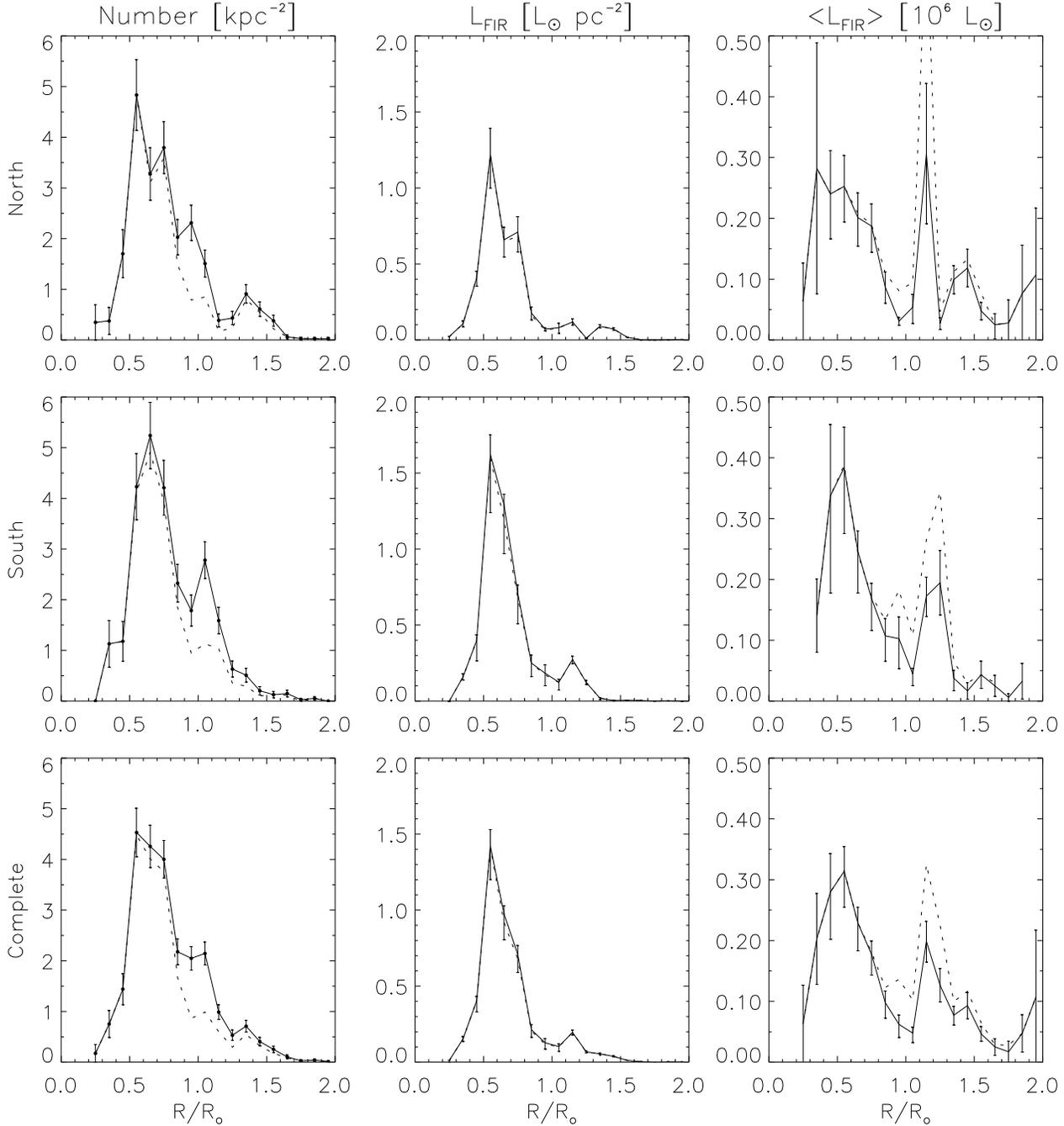}}
\end{center}
\caption[]{ The left column shows the number of detected sources divided by 
the sampled area, as a function of R, for the northern ($ 0^{\circ}\leq l\leq 180^{\circ}$) Galaxy (top row), the  southern 
($ 180^{\circ}\leq l \leq 360^{\circ}$) Galaxy (medium
row), and the whole Galaxy (bottom row). The center column shows, in the same
order, the azimuthally averaged face-on FIR galactic surface luminosity
produced by embedded massive stars; the right column shows the azimuthally
averaged FIR luminosity per embedded massive star forming region. The solid
line corresponds to the complete dataset. The dashed line represents the 
dataset corrected for incompleteness of the sample, as described in Sect. 3.5 } 
\end{figure*}     

%figure4
\begin{figure*}
\begin{center}
\resizebox{15cm}{!}{\epsfig{file=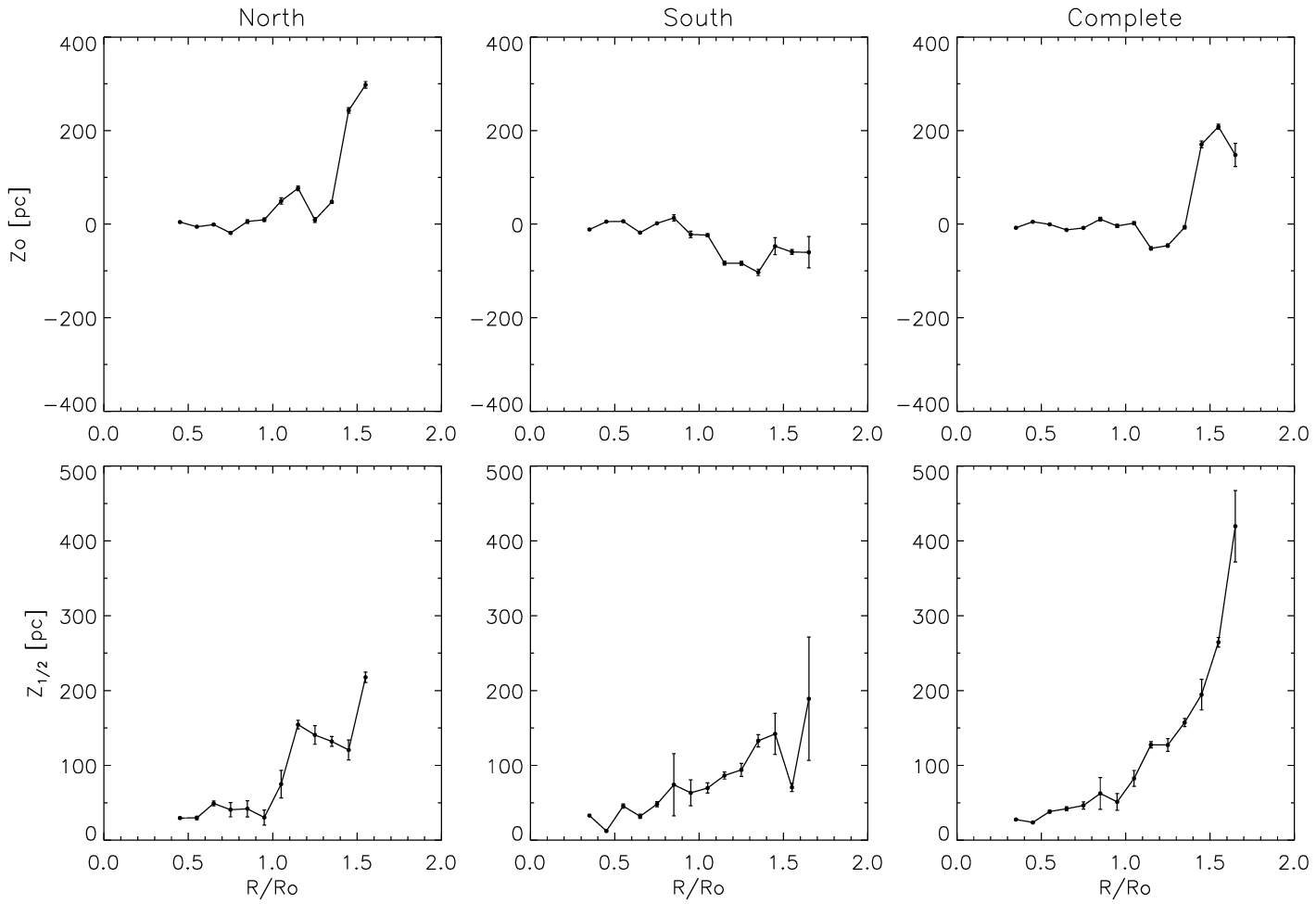}}
\end{center}
\caption[]{The azimuthally averaged centroid $Z_o$  (top panel) and the thickness 
(HWHM) $Z_{\frac{1}{2}}$ of the massive star formation layer (bottom panel) are 
shown  for the northern ($ 0^{\circ}\leq l\leq 180^{\circ}$) Galaxy (left); the  southern ($ 180^{\circ}\leq l \leq 360^{\circ}$) Galaxy (center); and the complete Galaxy (right)}
\end{figure*}

Using the kinematic information provided by the $\rm CS(2-1)$ line for the detected IRAS sources, we have derived their galactocentric distances assuming purely
circular motion about the galactic center. For simplicity we used a linear
rotation curve and the standard IAU constants, $R_{\circ} = 8.5$ kpc and
$V_{\odot} = 220$ km $\rm s^{-1}$. For lines of sight toward the galactic
center and anticenter the VLSRs of the sources are small and may have large
fractional errors due to systematic or random peculiar motions. Thus, for the
analysis here we have excluded all sources within 10$^{\circ}$ of the galactic
center and within 5$^{\circ}$ of the anticenter. Therefore, a total number of
748 detected sources have been used for the present analysis (Table 1), of
which 349 (47\%) are in the northern Galaxy (I and II quadrants) and 399 (53\%)
are in the southern Galaxy (III and IV quadrant). 

After finding the galactocentric distance $R$ for each source, they were binned 
in annuli $0.1\, R_{\circ}$ wide. The total number of sources in each ring was
divided by the ring sampled area to evaluate the azimuthally averaged number
surface density of massive star formation regions as a function of
galactocentric radius north, south, and for the complete dataset (Table 2; Fig.
3). The derivation depends only on the assumption of pure circular motion, and uses the kinematic galactocentric distance of each source, which can be derived one-to-one from the CS(2-1) velocity profiles. 

\subsection{Centroid and thickness of the OB star formation layer}

To derive the azimuthally averaged centroid $Z_o(R)$ and thickness 
$Z_{\frac{1}{2}}(R)$ (HWHM) of the massive star formation layer we
need to calculate the first and second order moments of the $Z$
distribution for each of the galactocentric rings analyzed. Within the solar
circle, however, for each source there are two possible kinematic heliocentric
distances allowed by the rotation curve and, therefore, two possible values
of $Z$. Following Paper I, we assume that the distribution
$P(Z)$ of sources is of gaussian form in $Z$,  

\begin{equation}
P(Z) = {\rm A}\,\exp{-{\frac{{\rm ln}2\,[Z - Z_o(R)]^2}{Z_{\frac{1}{2}}(R)^2}}}
\end{equation}
where $R$ is the galactocentric distance, $Z_o(R)$ and $Z_{\frac{1}{2}}(R)$ are fixed for each ring and A is a normalization constant. We then calculate the first and second order moments of the $Z$ distribution in each ring, $Z_1(R)$ and $Z_2(R)$, by evaluating a statistical average, using weights that depend on the near or far distance from the
plane, $Z_n$ and $Z_f$:

\begin{equation}
Z_1(R) ={\frac{1}{N(R)}}\sum_{j=1}^{N(R)} [ P_n(j)\, Z_n(j) +
P_f(j)\, Z_f(j)]  
\end{equation}
and 

\begin{equation}
Z_2(R)^2 = {\frac{1}{N(R)}}\sum_{j=1}^{N(R)} 
[ P_n(j)\, Z_n(j)^2 + P_f(j)\, Z_f(j)^2 ] 
\end{equation}
where $N(R)$ is the number of sources in a ring at galactocentric distance 
$R$, $P_n = P(Z_n)$, $P_f = P(Z_f)$, and the normalization
condition $P_n + P_f = 1$ is used. 

The centroid $Z_o(R)$ is equal to the first order moment
$Z_1(R)$, and the thickness $Z_{\frac{1}{2}}(R)$ (HWHM) is given by

\begin{equation}
Z_{\frac{1}{2}}(R)^2 = 2 \, {\rm ln}2 \,[Z_2(R)^2 - Z_o(R)^2]
\end{equation}

The weights $P_n$ and $P_f$ depend in turn of $Z_o$ and $
Z_{\frac{1}{2}}$ at each ring, so the problem is cyclic. It can be solved by
an iterative procedure (Casassus 1995), starting with reasonable values, like
$Z_o = 0$ and $Z_{\frac{1}{2}} = 60$ pc, close to the average
HWHM of the H$_2$ layer within the solar circle. 

While it is clear that each source is either at the near or at the far
distance within the solar circle, the weights used in the statistical
averages account for the probability of each source to be at the near or
far distance by using the well known {\it latitude effect}, i. e., the
farther away a source, the more likely to be close to the galactic plane. The
weights are used only to derive large scale averages in a statistical sense, and
not individually to derive a mean distance for each source. The results of the 
iterative calculation are displayed in Fig. 4 and listed in Table 2; empty
spaces in the table correspond to those rings where the number of detected
sources is less than 3, not enough to have a convergent iteration so as to derive values for
$Z_o$  and $Z_{\frac{1}{2}}$.    

\subsection{Azimuthally averaged FIR surface luminosity produced by embedded
OB stars} 

The azimuthally averaged face-on FIR surface luminosity is evaluated by summing 
the luminosities of the sources in each ring and dividing the result by the
sampled area. The average luminosity per source is the ratio of the summed
luminosity in each ring to the number of sources considered. We estimate the FIR flux of a source as the sum over the four IRAS bands,

\begin{equation}
\label{eq:Firas}
F_{IRAS}=\sum_{j=1}^{4} \nu F_{\nu}(j),
\end{equation}
where $F_{\nu}(j)$ are the IRAS band flux densities, as listed in the
IRAS PSC (P\'erault 1987). 
The FIR luminosity for each source, outside the solar circle, is computed as $L\,=\,4\pi\,D^2\,F_{IRAS}$, where D is its kinematic distance, obtained from its CS(2-1) line profile. We overcome the two-fold distance ambiguity, within the solar circle, by evaluating the total luminosity in each ring, $L(R)$, as a weighted sum. Like for the $Z$ moments, in that sum we use the {\it latitude effect} to take into account the probability for each source to be at the near or far distance in the galactic disk, i. e.,  

\begin{equation}
L(R) = \sum_{j=1}^{N(R)} [ P_n(j)\, L_n(j) + P_f(j)\, L_f(j)]
\end{equation}
where $L_n\,=\,4\pi\,D_n^2\,F_{IRAS}$ and  $L_f\,=\,4\pi\,D_f^2\,F_{IRAS}$ are the luminosities that the source would have at the  near and far distances, respectively. The weights $P_n = P(Z_n)$ and $P_f = P(Z_f)$ are obtained using the values of $Z_o$  and $Z_{\frac{1}{2}}$ listed in Table 2. For the innermost ring, at $R/R_{\circ}$ = 0.25, where no model is available, and at $R/R_{\circ}$ = 0.35 in the north, we use the subcentral distances to estimate the luminosities. The FIR surface luminosity produced by regions of massive star formation and the average FIR luminosity per region, as a function of $R$, are shown in Table 3 and Fig. 3. The procedure used to remove the distance ambiguity, within the solar circle, is fully equivalent to the {\it direct partitioning} technique used to derive the  H$_2$ surface density in Paper I
(see Eq. (4) and Fig. 15).

%table3
\begin{table}
\
\caption{FIR surface luminosity}
\[
\begin{array}{l@{\hspace{1cm}}l@{\hspace{1cm}}l@{\hspace{1cm}}l}
\hline
\noalign{\smallskip}
R_{\rm bin} & L_{\rm FIR(North)} & L_{\rm FIR(South)} & L_{\rm FIR(Comb)}\\
(R_{\rm o}) & (L_{\odot}\,{\rm pc}^{-2}) & (L_{\odot}\,{\rm pc}^{-2}) &
(L_{\odot}\,{\rm pc}^{-2})\\
\noalign{\smallskip}
\hline
\noalign{\smallskip}
      0.25 & 4.60E-02 & 0.00E+00 & 2.30E-02\\
      0.35 & 2.35E-01 & 1.59E-01 & 1.80E-01\\
      0.45 & 4.09E-01 & 3.98E-01 & 4.03E-01\\
      0.55 & 1.22E+00 & 1.62E+00 & 1.42E+00\\
      0.65 & 6.59E-01 & 1.29E+00 & 9.75E-01\\
      0.75 & 7.10E-01 & 7.09E-01 & 7.10E-01\\
      0.85 & 1.77E-01 & 2.50E-01 & 2.13E-01\\
      0.95 & 7.15E-02 & 1.83E-01 & 1.27E-01\\
      1.05 & 8.11E-02 & 1.24E-01 & 1.03E-01\\
      1.15 & 1.18E-01 & 2.74E-01 & 1.96E-01\\
      1.25 & 1.18E-02 & 1.23E-01 & 6.72E-02\\
      1.35 & 9.03E-02 & 1.92E-02 & 5.47E-02\\
      1.45 & 7.22E-02 & 3.39E-03 & 3.78E-02\\
      1.55 & 1.81E-02 & 5.47E-03 & 1.18E-02\\
      1.65 & 1.49E-03 & 3.74E-03 & 2.62E-03\\
      1.75 & 7.80E-04 & 1.67E-04 & 4.74E-04\\
      1.85 & 2.02E-03 & 1.72E-03 & 1.87E-03\\
      1.95 & 2.68E-03 & 0.00E+00 & 1.34E-03\\
     \noalign{\smallskip}
     \hline
\end{array}
\]
\end{table}

\subsection{Errors derivation}  

The errors in the number surface density $N$ consider only the counting 
statistics assuming a Poisson distribution. The uncertainties in the best
values of the parameters $Z_o$  and $Z_{\frac{1}{2}}$ were estimated  on the
basis that the errors are dominated by the peculiar motions of the sources,
which we assume to be of $\sim 5$ km s$^{-1}$  (VLSR). Using this value we have
then calculated the uncertainties in the kinematic distances, and in the
quantities which are derived from the kinematic distances. At $1.1 \leq R/R_{\circ} \leq 1.2 $, in the northern Galaxy, the average luminosity per source is dominated by one source, G70.293 (BNM), a very bright UC H II region in the Perseus spiral arm. 

The uncertainties in the $L_{FIR}$ values have also a contribution from the
instrumental errors in the IRAS detectors. We used the errors in the 100 $\mu$m 
band as representative of the fractional flux uncertainty for each source (Casassus et al. 2000). When the flux value  obtained by IRAS is only an upper limit, the error has been
assumed to be of $-100$\%. There is also some level of correlation in the
position of the sources; as an example, for massive star forming GMCs 
in the IV galactic quadrant there are $\sim 4$ 
embedded UC H II regions per GMC (Bronfman 1992). We have taken into
account such correlation by computing for each galactocentric ring $N_{\mathrm corr}(R)$, the average number of sources that lie within a box of sides 2$^{\circ}$ by 2$^{\circ}$ by 20 km s$^{-1}$ - a representative maximum value for a GMC - and multiplying the errors in $L_{FIR}$ at each ring by $\sqrt{N_{\mathrm corr}(R)}$.

\subsection{Spiral structure and the axisymmetric analysis}

What are the shortcomings in our analysis caused by deviations from
axial symmetry in the galactic distribution of embedded OB stars? Images of
external galaxies clearly show a high concentration of massive star formation
in spiral arms, and the same should be expected for our own Galaxy. While the
large scale distribution of OB stars in the Milky Way cannot be traced optically
because of dust obscuration in the disk, H II regions produced by these stars
can be observed in radio-recombination lines, like H\,109$\alpha$, and have been
used to study the spiral arm structure of our Galaxy (Georgelin \& Georgelin
1976; Downes et al. 1980). The UC H II regions, which point at very young star 
forming regions, still embedded in their parental molecular clouds, could be
even more confined to spiral arms than the extended H II regions observed in
centimetric wavelengths. 

%figure5
\begin{figure}
\begin{center}
\resizebox{9cm}{!}{\epsfig{file=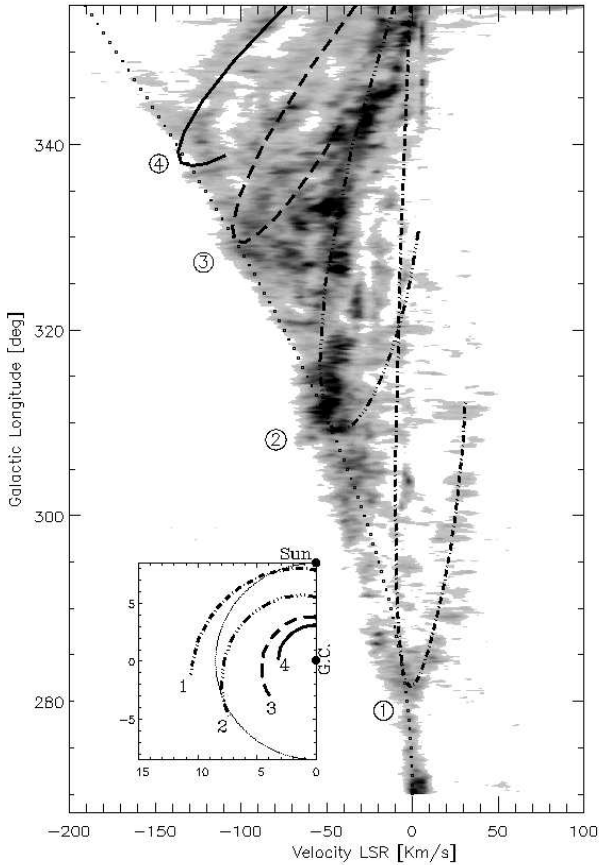}}
\end{center} 
\caption[]{ The spiral arm pattern used for the analysis. The arms are labelled
1 (Carina), 2 (Crux), 3 (Norma), and 4 (3-kpc). The CO emission (Paper I; 
Grabelsky et al. 1987) integrated in 
latitude from $-2\degr$ to $2\degr$ is represented in gray-scale. Lowest 
contour level is at 3 $\sigma$. The scale in the inset figure is in kpc, and 
the galactic center is labelled G.C. } 
\end{figure}

Spiral arms within the solar circle appear to be more separated and less 
difficult to trace in the southern Galaxy than in its northern counterpart.
These arms, as outlined by CO emission from molecular clouds, are 
tangent to the line of sight at longitudes $283\degr$ (Carina); $308\degr$ (Crux);
$328\degr$ (Norma); and $337\degr$ (3-kpc expanding arm). The tangent
longitudes are traced by discontinuities in (a) the galactic CO emission
integrated in velocity and latitude I($l$) (Paper I; Grabelsky et al. 1987) and in (b) the rotation curve derived from CO observations (Alvarez et
al. 1990). The distribution of H II regions in the southern Galaxy, derived
from an extensive survey of hydrogen recombination lines toward 5 GHz continuum
sources (Caswell \& Haynes 1987), is consistent also with the presence of these four spiral arm segments in the IV galactic quadrant. 

To test our axisymmetric analysis we have derived the mean radial distribution
of UC H II regions in the southern Galaxy in the framework of this generally
accepted spiral arm model. Most GMCs identified through CO observations in the southern Galaxy are coincident, in space and velocity, with one or more radio H II regions from the Caswell \& Haynes (1987) sample, as well as with several UC H II regions from the dataset we analyze here. We have used those radio H II regions from Caswell \& Haynes that have well defined distances, derived through H$_2$CO absorption lines, to remove the distance ambiguity for the associated GMCs and for their embedded UC H II regions (Bronfman 1992). These GMCs, together with the tangent points listed above, trace a spiral arm pattern in the longitude-velocity diagram (Fig. 5). The distance ambiguity for the remaining GMCs is removed by assigning each one to the closest spiral feature in the longitude-velocity diagram. For some GMC's in the far side of the Crux arm, close to the solar circle, we use their size-to-linewidth ratio to distinguish them from local clouds with small VLSRs.       

The mean radial distribution of FIR surface luminosity produced by embedded
UC H II regions, obtained using the spiral model, is fairly similar to that
obtained using the axisymmetric analysis (Fig. 6a). The average luminosity
per source (Fig. 6b) shows a secondary peak, close to $R = R_o$, caused by the
far side of the Crux spiral arm cutting through the solar circle at $316\degr \leq l 
\leq 320\degr$. We do not quote errors in this comparison because
the spiral model, although well updated, involves much uncertainty and we are
mainly interested in the general trend of the mean radial variations. It is apparent that the peaks are at the same positions and the general shapes are the
same for both analyses.  The axisymmetric analysis, however, seems to 
overestimate the mean FIR luminosities for radii other than the main peak locus.

%figure6
\begin{figure}
\begin{center}
\resizebox{9cm}{!}{\epsfig{file=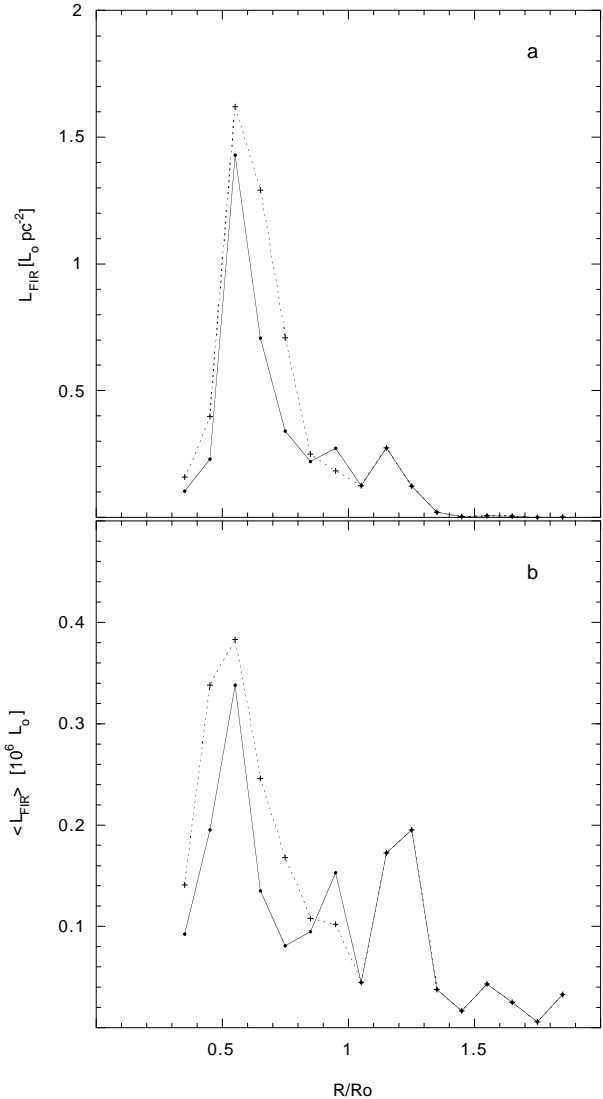}}
\end{center} 
\caption[]{ (a) FIR surface luminosity and (b) average luminosity per source, as a function of R, for the southern Galaxy. The solid line shows the results obtained by
removing the two-fold distance ambiguity, within the solar circle, using a
spiral model. The dotted line shows the results obtained using the axisymmetric analysis} 
\end{figure}

\subsection{Completeness of the sample and contamination by low luminosity 
sources in the solar neighborhood}

The question of completeness has been adressed in general by WC89, who show 
that UC H II regions are tightly confined
in the FIR color-color plot, and can be easily distinguished from other
entries in the IRAS PSC. Thus, the distinctive FIR colors of embedded massive
stars can be used to count the number of such objects in the Galaxy. They 
conclude that all embedded O stars and some B stars will be detected by the 
color-color criterion applied to the IRAS PSC in the Galaxy. Contamination of 
the sample by cloud cores with lower mass stars has been analyzed by 
Ramesh \& Sridharan (1997), who argue that the total number of potential UC 
H II regions in the WC89 sample is only about 25\% of the sample. Our own criterion,
based on the detection of the CS(2-1) line toward each of the UC H II region candidates (BNM), brings down the original WC89 sample by a factor of $\sim 2$.   

%figure7
\begin{figure}
\begin{center}
\resizebox{9cm}{!}{\epsfig{file=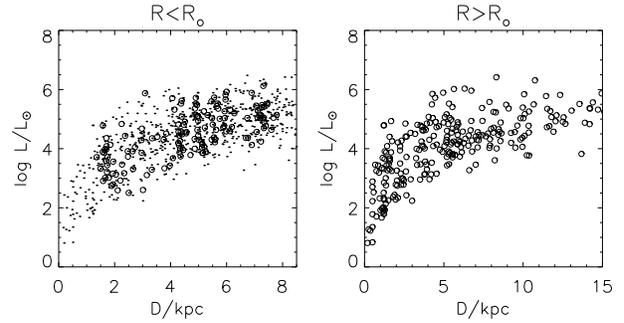}}
\end{center}
\caption[]{ FIR luminosity versus distance for all the sources in the sample. 
Within the solar circle (left) the dots represent distances and
luminosities obtained through the weighting method (See Sect. 3), while the circles represent subcentral sources, with better determined distances. Outside the solar circle (right) all distances are well defined} 
\end{figure}

%figure8
\begin{figure}
\begin{center}
\resizebox{9cm}{!}{\epsfig{file=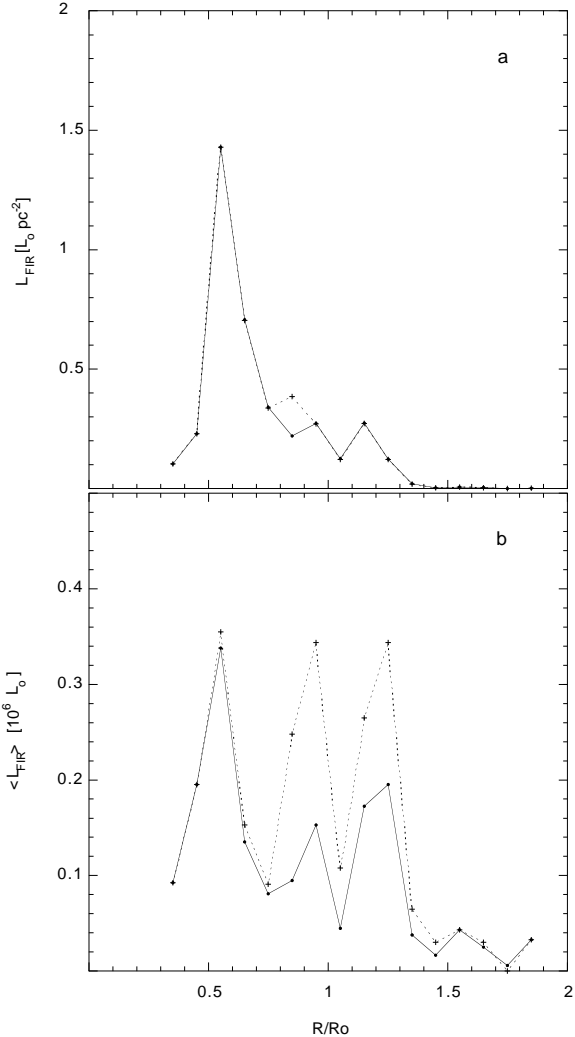}}
\end{center}
\caption[]{ (a) FIR surface luminosity and (b) average luminosity per source, as a function of (R), in the southern Galaxy. The two-fold distance ambiguity, within the solar
circle, has been removed using a spiral model. The
results obtained using the complete dataset are shown with the solid line; 
results obtained when leaving out all sources with FIR
luminosities lower than 8000 $L_{\odot}$ are shown by the dotted line} 
\end{figure}

We analyze here the question of completeness of the sample as a function of
distance; this is particularly necessary to understand the behaviour of the
number density and mean luminosity of sources near the solar circle. The FIR
luminosity versus distance for all sources in the sample analyzed is shown in 
Fig. 7.  The fairly homogeneous distribution in distance of sources with FIR luminosities larger than ${\rm log} (L/L_{\odot}) \sim 4 $ suggests that our sample is complete above such luminosity level.  Closer than $\sim 3$  kpc from the Sun the sample appears to be  contaminated by lower luminosity sources. To test our results we have reanalyzed our data applying a lower FIR luminosity cutoff to the sample, both for the spiral and for the axisymmetric model. Wouterlout et al. (1995, their Fig. 20b) show that the WC89 FIR color criterion for UC H II regions tends to select point sources with ${\rm log}(L/L_{\odot}) \ge 4 $. Thus, even if the IRAS PSC were more sensitive, there would be little point in reducing the luminosity limit much under $10^4 L_{\odot}$.   

The CS(2-1) detection requirement appears not to introduce additional biases so we
consider unnecessary to address separately the completeness of the CS survey. The observed antenna temperatures T$_{\rm A}$ within the solar circle, in particular for the subcentral sources, are not distance dependant. For sources outside the solar circle, the average T$_{\rm A}$ is 0.9 K for distances between 2 and 6 kpc, and 0.6 K beyond.  The SEST beamwidth of 50" (FWHM) projects a linear size of 2.4 pc at 10 kpc, about the typical size of CS cores harbouring massive star formation (Bronfman 1992), so beam-dilution is not a stringent problem for the CS detections.       
 
The mean radial distribution of the FIR surface luminosity does not change
noticeably when a lower cutoff luminosity is applied to the sample.  We have
computed the azimuthally averaged FIR surface luminosity and the mean
luminosity per source as a function of galactocentric radius, using the same spiral analysis as in Sect. 3.4 within the solar circle, and a lower cutoff luminosity $ {\rm log} (L/L_{\odot})= 3.9$. The results are presented in Fig. 8, and the number of sources above the cutoff is listed, for each ring, in Table 1 (in parenthesis). At the cutoff value used, the three peaks in the average luminosity per source have the same magnitude, i.e $L_{\rm FIR} \sim 0.35 \, 10^6 L_{\odot}$. These peaks are dominated by sources located in the far sides of the Norma arm ($R = 0.55\, R_{\circ}$), the Crux arm ($R = 0.95 \,R_{\circ}$), and the Carina arm ($R = 1.25\, R_{\circ}$). 

While we cannot, for the time being, confirm that the average luminosity per source diminishes with galactocentric radius, it is tempting to suggest that UC H II regions appear to be very good tracers of spiral arm structure in the Galaxy, and that a full spiral analysis of their distribution in the Milky Way is mandatory. While in the present paper we perform a zero'th order axisymmetric comparison of the distribution of OB star formation regions and of molecular clouds, there is an evident necessity of resolving the two-fold distance ambiguity for all these regions in order to refine the present results, within the solar circle, and extend them toward a more exact analysis of the mean luminosity per source, which has a direct impact on the determination of the IMF for massive star forming regions.

The FIR surface luminosity derived from the axisymmetric analysis, our main result,  remains also fairly unchanged (Fig. 3) when a lower luminosity cutoff is applied to the sample. The FIR surface luminosity at each radius is the product of the number of sources considered, per unit area, times their average luminosity. While the number surface density of sources decreases when a luminosity cutoff is applied, particularly for the solar neighborhood, the average luminosity per source increases, since we  are not taking into account the less luminous sources for the average. Therefore the results we present next, based on our determination of the FIR surface luminosity produced by embedded massive stars and its comparison with the H$_2$ surface density within the solar circle, appear to be not too affected by problems of completeness. As for the effects of spiral structure over the derived FIR surface density, they are small and would have presummably the same effects over the distribution of molecular hydrogen derived using an axisymmetric model. 

%figure9
\begin{figure*}
\begin{center}
\resizebox{15cm}{!}{\epsfig{file=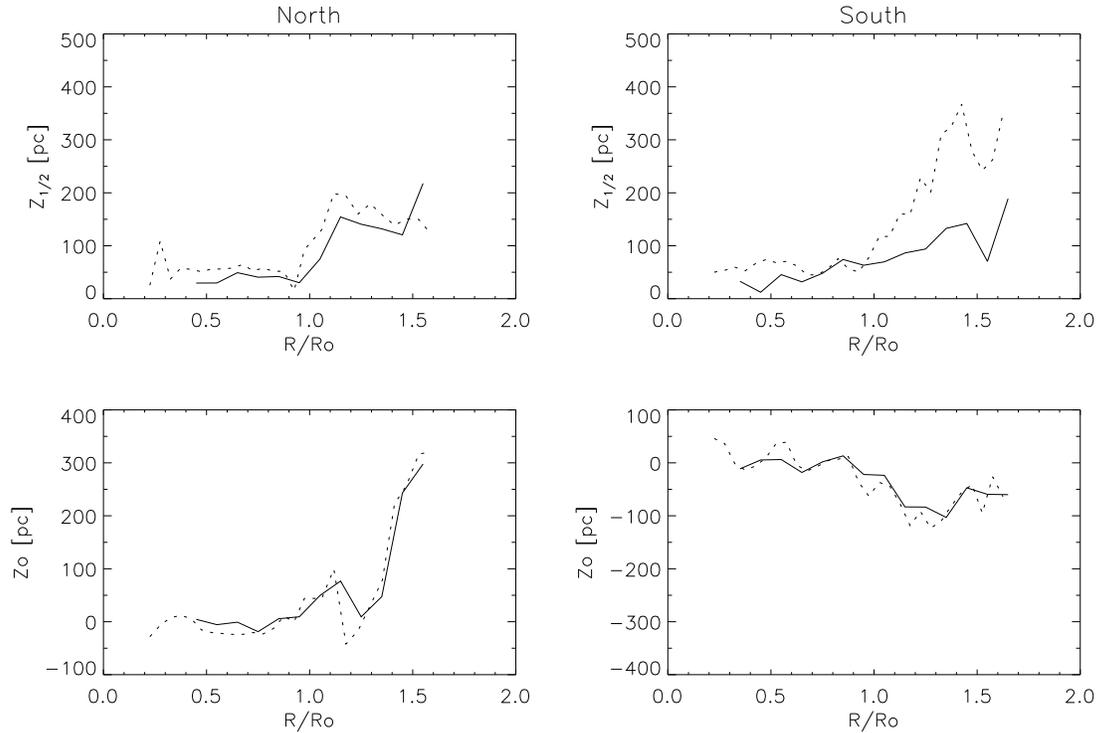}}
\end{center}
\caption[]{ The azimuthally averaged thickness (HWHM) $Z_{\frac{1}{2}}$ (top) and the centroid $Z_o$ (bottom) of the massive star formation layer (solid line), as a function of R, are compared to those of H$_2$ (dashed line), as derived from the Columbia U. - U. Chile deep CO surveys of the Galaxy} 
\end{figure*}

%figure10
\begin{figure}
\begin{center}
\resizebox{8.5cm}{!}{\epsfig{file=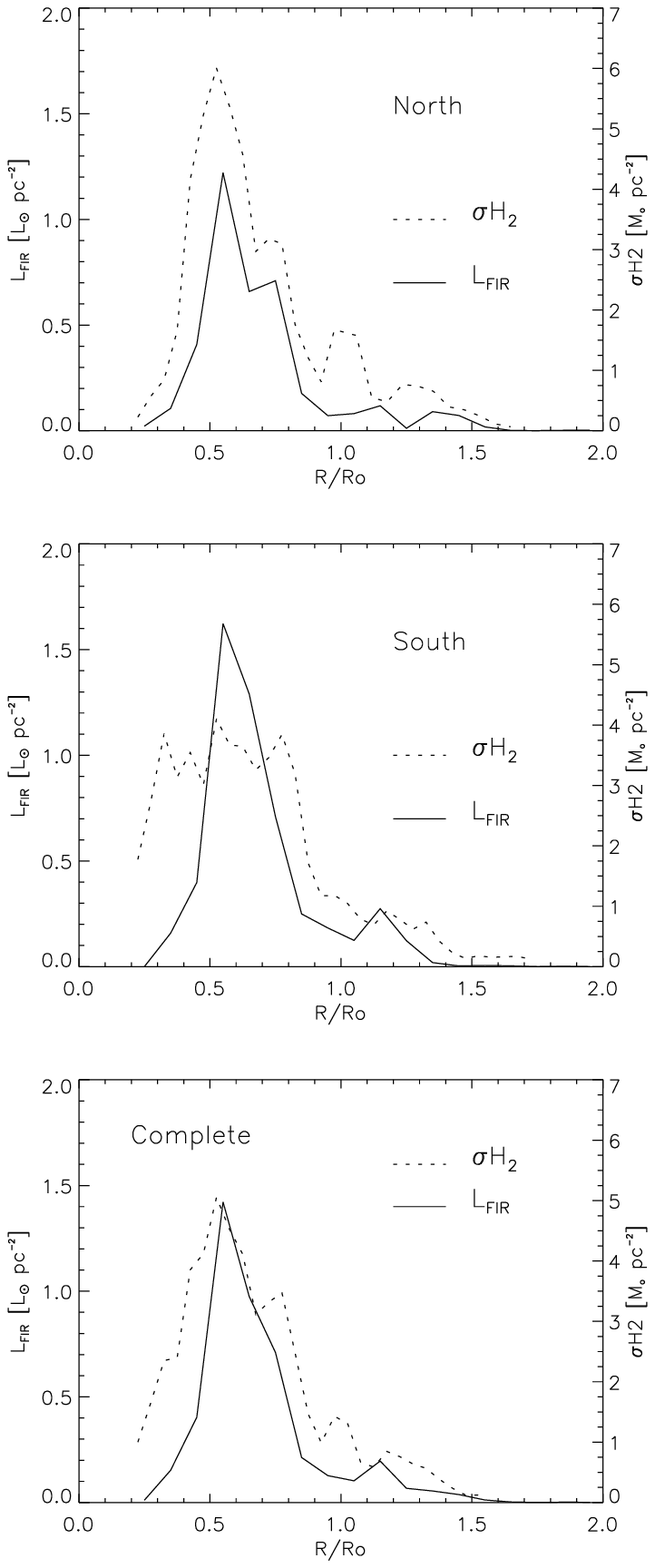}}
\end{center}
\caption[]{From top to bottom: azimuthally averaged face-on FIR surface 
luminosity, as a function of R, originated by embedded massive 
stars (solid line) for the northern Galaxy ($0^{\circ}\leq l \leq 180^{\circ}$); 
the southern Galaxy ($ 180^{\circ}\leq l \leq 360^{\circ}$); and the whole Galaxy, compared with molecular hydrogen face-on surface density  (short dashed line), as derived from the Columbia U. - U. Chile deep CO surveys of the Galaxy. Data are uncorrected for the expansion velocity field at the 3-kpc arm locus (Paper I)} 
\end{figure} 

\section
{The galactic disk traced by embedded OB stars and by molecular clouds}

\subsection
{Centroid and thickness of the disk}

The formation of OB stars in the galactic disk is distributed in a layer 
coincident but thinner than that of molecular clouds as traced by CO. Fig. 9
shows the radial variations of the centroid $Z_o$  and the thickness (HWHM)
$Z_{\frac{1}{2}}$ of the embedded massive star layer for the southern and
northern Galaxy, compared with their H$_2$ counterparts.  The molecular gas parameters were derived from the Columbia U. - U. Chile deep CO surveys of the I and IV galactic quadrants ( Paper I; Cohen et al. 1986); from surveys of the Carina region by Grabelsky et al. (1987) and of the III galactic quadrant by May et al. (1988, 1993), analyzed by Casassus (1995); and from  a {\it composite} CO dataset of the II quadrant analyzed by Digel (1991). 

The average thickness of the embedded massive star distribution, within the solar circle, is 73 pc (FWHM), compared with 118 pc (FWHM) for the layer of molecular gas traced by CO (Tables 6 and 7). It has been shown by Dame \& Thaddeus (1994), however, that there are possibly two components for the molecular gas layer: a thin one, with a FWHM of 80 pc, similar to the thickness of the young massive stars layer derived here, and a thick and diffuse one, with a FWHM in excess of 500 pc.  

The warping of the disk outside the solar circle, as traced by the centroid $Z_o$ of both the massive star formation and the H$_2$ layer, appears to be more pronounced in the northern Galaxy than in the south. A tentative  explanation is that the spiral arms outside the solar circle can be traced within a larger longitude range in the south than in the north; we may be therefore  averaging out the warp in azimuth in the south because of the large longitude span (Casassus 1995; May et al. 1997). For example, it is apparent from Fig. 2 than there are many more sources detected outside the solar circle in the IV quadrant than in the I quadrant of the Galaxy. 

The flaring of the molecular gas layer in the outer Galaxy is evidenced by both the molecular and by the embedded massive star distribution in the south. However, while the thickness of the massive star layer in the northern outer Galaxy grows like that in the south, the mean thickness of the molecular hydrogen layer in the northern outer Galaxy is only about half that of its  southern counterpart.  A possible explanation is that while the longitude coverage of our survey of embedded massive stars is equally complete north and south, the {\it composite} CO data set of the northern outer Galaxy analyzed by Digel (1991)  only covers the region $65\degr \, \leq l \leq 116\degr$, compared to the complete longitude coverage of the III and IV quadrant CO surveys used by Casassus (1995) in his analysis. Therefore the azimuthal range for the average is much larger in the south than in the north; averaging the galactic warp over large scales would increase the mean thickness derived for the molecular gas layer.  

\subsection{Embedded OB stars, FIR surface luminosity, and H$_2$ surface density}

The mean radial distribution for the 748 massive star forming regions analyzed here (Fig 3) is maximum at $R \sim 0.55 \,R_{\circ}$, falls sharply toward the galactic center, and decays exponentially toward the outer Galaxy  (Fig. 10). Of these regions, 492 ($67\%$) are found at $R\,\leq\,R_{\circ}$, and 332 ($44\%$) 
at $0.4 \, \leq R/R_{\circ} \,\leq\, 0.8 $, the region of the molecular annulus (Paper I). The fraction of $44\%$ grows to $55\%$ when the contamination of the sample by low luminosity regions in the solar neighborhood is considered (Sect. 3.5). A similar annular radial dependence for the number surface density of OB associations in the inner Galaxy has been found by Mc Kee \& Williams (1997) using an older compilation of radio H II regions (Smith et al. 1978) binned in rings 2 kpc wide. Such annular shape was found also by  Kent et al. (1994) for the NIR emission attributed to M giants, based on large scale satellite observations.

The FIR surface luminosity produced by embedded massive stars for the Galaxy as a whole is even more restricted to the molecular annulus than their number density. Integration over the whole galactic disk yields a total FIR luminosity of $1.39 \, 10^{8}\, L_{\odot}$. Within the solar circle the FIR luminosity is $1.13 \, 10^{8}\, L_{\odot}$, $ 81$\% of the total, and for $0.4 \leq R/R_{\circ} \leq  0.8 $, the FIR luminosity is $0.97 \, 10^{8}\, L_{\odot}$, representing $ 70\%$ of the total. The radial dependance of the FIR surface luminosity produced by embedded massive stars has also a narrower maximum than the H$_2$ surface density. A gaussian fit to the azimuthally averaged FIR surface luminosity, within the solar circle, yields a FWHM of $0.28 \,R_{\circ}$ (Table 4), $\sim55$\% the FWHM obtained for the H$_2$ radial distribution, $0.51 \,R_{\circ}$ (Table 5). Down from the maximum, toward the outer Galaxy, the FIR surface luminosity decreases, as a function of R, much faster than the H$_2$ surface density, with an exponential scale length of $0.21 \,R_{\circ}$, compared to $0.34 \,R_{\circ}$ for the H$_2$.  
 
The ratio of FIR luminosity produced by embedded OB stars to H$_2$ mass, which may be used as a large scale tracer of massive star formation efficiency, is not constant throughout the Galaxy and has also a peak at $R\,\sim0.55\,R_{\circ}$. Integrating the H$_2$ surface density, as traced by the Columbia U. - U. de Chile CO surveys of the Galaxy, we compute a total H$_2$ mass of $ 6.1  \, 10^{8}\, M_{\odot}$ for $0.2 \, \leq R/R_{\circ} \,\leq\, 1 $, and of $ \sim\,2\, 10^{8}\, M_{\odot}$ for 
$ 1 \, \le R/R_{\circ} \, \leq 1.7$. For all the CO data we have used
main beam antenna temperatures $T_{\rm mb}$, with a main beam efficiency of 0.82 
(Paper I); CO to H$_2$ conversion factor $N({\rm H}_2)/W_{\rm CO} = 1.56\,10^{20}\,{\rm molecules}\,{\rm cm}^{-2}\,({\rm K}\,{\rm km}\,{\rm s}^{-1})^{-1}$ derived using the same CO data and EGRET gamma-ray observations by Hunter et al. (1997); a distance of the Sun to the galactic center of $ R_{\circ} = 8.5 \,{\rm kpc}$; and no correction for helium abundance. The average FIR luminosity to H$_2$ mass ratio, within the solar circle, is $\sim 0.18 \,L_{\odot}/M_{\odot}$; for $0.4 \leq R/R_{\circ} \leq 0.8 $ the ratio grows to $\sim 0.23\, L_{\odot}/M_{\odot}$; and for $0.5 \leq R/R_{\circ} \leq 0.6$ there is a maximum of $\sim 0.29\, L_{\odot}/M_{\odot}$. Outside the solar circle, for $R/R_{\circ} \leq 1.7$, the derived mean ratio is $\sim 0.13\, L_{\odot}/M_{\odot}$; this figure can be considered as an upper limit, since the conversion factor $N({\rm H}_2)/W_{\rm CO}$ has been argued to be from 2 to 4 times larger in the outer Galaxy (May et al. 1997).       

The radial distribution for the complete set of OB star formation candidate sites  from WC89 has been derived by Comer\'on \& Torra (1996), assuming axial symmetry but without the use of any kinematic information. Our present results do not confirm the peak in massive star formation that Comer\'on \& Torra (1996) find at $R = 2$ kpc, although it is fair to say that our analysis is oriented to the galactic disk, and that we have therefore cut out the region within $10\degr$ of the galactic center. However, we detect only 1 source in the range $1.70\,{\rm kpc} \leq R \leq 2.55\,{\rm kpc}$, where we sampled 50\% of the face-on area, while for their result to hold there should be more than 50 OB star forming regions in the area we sampled. 

\subsection
{Large scale asymmetries within the solar circle}

%Table4
\begin{table*}
\
\caption{Azimuthally averaged face-on FIR surface luminosity produced by regions of massive star formation}
\
\begin{tabular}{l@{\hspace{0.75cm}}l@{\hspace{0.5cm}}l@{\hspace{1.25cm}}l@
{\hspace{1.0cm}}l}
\hline
\noalign{\smallskip}
Sampled region & Gaussian maximum$^{\rm a}$ & Gaussian center$^{\rm
a}$ & Gaussian FWHM$^{\rm a}$ & Exponential scale-length$^{\rm b}$\\ 
	& ($L_{\odot}\,{\rm pc}^{-2}$) & ($R_{\circ}$) & ($R_{\circ}$) &
($R_{\circ}$)\\ 
\noalign{\smallskip}
\hline
\noalign{\smallskip}
North & $1.02\pm 0.16$ & $0.60\pm 0.02$ & $0.31\pm 0.06$ &
$0.21\pm 0.03$\\
South & $1.59\pm 0.16$ & $0.60\pm 0.01$ & $0.26 \pm 0.03$ & $0.22 \pm 0.02$\\
Complete & $1.29\pm 0.15$ & $0.60\pm 0.02$ & $0.28\pm 0.04$ & $0.21\pm 0.02$\\
\noalign{\smallskip}
\hline
\end{tabular}
\begin{list}{}{}
\item[$^{\rm a}$] Fit between $0.2\,R_{\circ} \leq R \leq R_{\circ}$
\item[$^{\rm b}$] Fit between $0.5\,R_{\circ} \leq R \leq 2\,R_{\circ}$  
\end{list}
\end{table*}

%Table5
\begin{table*}
\
\caption{Azimuthally averaged face-on surface density of molecular hydrogen}
\
\begin{tabular}{l@{\hspace{0.75cm}}l@{\hspace{0.5cm}}l@{\hspace{1.25cm}}l@
{\hspace{1.0cm}}l}
\hline
\noalign{\smallskip}
Sampled region & Gaussian maximum$^{\rm a}$ & Gaussian center$^{\rm
a}$ & Gaussian FWHM$^{\rm a}$ & Exponential scale-length$^{\rm b}$\\ 
& ($M_{\odot}\,{\rm pc}^{-2}$)  & ($R_{\circ}$) & ($R_{\circ}$) & ($R_{\circ}$)\\ 
\noalign{\smallskip}
\hline
\noalign{\smallskip}
North & $5.5\pm 0.4$ & $0.57\pm 0.01$ & $0.37\pm 0.03$ & $0.29\pm 0.02$\\
South & $3.9\pm 0.2$ & $0.55\pm 0.02$ & $0.71 \pm 0.08$ & $0.40\pm 0.04$\\
Complete & $4.5\pm 0.2$ & $0.57\pm 0.01$ & $0.52\pm 0.03$ & $0.34\pm 0.02$\\
\noalign{\smallskip}
\hline
\end{tabular}
\begin{list}{}{}
\item[$^{\rm a}$] Fit between $0.2\,R_{\circ} \leq R \leq R_{\circ}$
\item[$^{\rm b}$] Fit between $0.5\,R_{\circ} \leq R \leq 2\,R_{\circ}$  
\end{list}
\end{table*}

Among the most striking results of the early CO surveys of the Galaxy were the 
large scale north-south deviations from axial symmetry obtained from separate
fits of axisymmetric models to the CO data (Robinson et al. 1983; Cohen et al.
1985). The maximum molecular gas face-on surface density in the molecular
annulus is a factor of 1.47 higher in the I galactic quadrant than in the IV
quadrant (Paper I), where the maximum is really a plateau
extending roughly for $ 0.4 \leq R/R_{\circ}  \leq 0.8 $.
Because massive star formation occurs mostly in giant molecular clouds, it would 
be expected to find a similar trend in the FIR surface luminosity produced by embedded massive stars. What we have found, however, is an {\it enhancement} of massive star formation in the southern Galaxy within the solar circle.

A fraction of 58\% of the FIR luminosity produced by embedded massive stars, within the solar circle, comes from the southern Galaxy - while the total H$_2$ mass, as traced by CO, is about the same north and south. A possible explanation of such result could be that massive stars in the I galactic quadrant are  preferentially in the far side of the Galaxy, and/or that in the IV quadrant they are preferentially in the near side. Thus, by assuming axial symmetry in our models, we could be artificially inflating the  luminosity in the south or lowering it in the north. However, because the FIR luminous dense molecular gas cores we have sampled in the CS(2-1) are well correlated in space and velocity with the population of giant molecular clouds in the Galaxy as traced by CO (Bronfman 1992), the same asymmetry should be found in the distribution of molecular gas. But what we
find here for the FIR surface luminosity distribution is an asymmetry from
north to south which is opposite in sign to that of the molecular gas surface
density.  

Massive star formation per unit H$_2$ mass, as a consequence, is higher in the IV than in the I galactic quadrant. The average FIR luminosity to H$_2$ mass ratio in the southern Galaxy, for $0.2 \leq R/R_{\circ} \leq 1$, is $\sim 0.21 \,  L_{\odot}/M_{\odot}$; for $0.4 \leq R/R_{\circ} \leq 0.8 $ grows to 0.28; and for $0.5 \leq R/R_{\circ} \leq 0.6 $  we find a galactic maximum of $\sim 0.41\, L_{\odot}/M_{\odot}$. In comparison, the  corresponding numbers in the northern Galaxy are $\sim 0.16\, L_{\odot}/M_{\odot}$, $\sim 0.18\, L_{\odot}/M_{\odot}$, and $\sim 0.21\, L_{\odot}/M_{\odot}$. The enhancement of massive star formation at $R \sim 0.55\,R_{\circ}$ in the southern Galaxy appears to be correlated with the position of the Norma spiral arm tangent, where a large fraction of the most luminous sources in the sample are found. Massive star formation enhancements of this kind have been observed in the spiral arms of external galaxies, like for instance in M51 (Nakai et al. 1994), although there the starburst seems to be driven by a companion galaxy. In our own galaxy such burst could be driven by a bar - that would have to be large enough to produce an effect at almost 5 kpc from the galactic center.

\subsection{The contribution of embedded OB stars to the FIR luminosity 
of GMCs in the Galaxy}

Preliminary results of our ongoing analysis of GMCs harbouring embedded UC H II regions in the southern Galaxy show a fairly linear relation between the FIR luminosity from embedded OB stars and the virial mass of the clouds, with a mean FIR luminosity to mass ratio of $\sim 0.4 \, L_{\odot}/M_{\odot}$ (Bronfman et al. 2000). This ratio grows to $\sim 4 \, L_{\odot}/M_{\odot}$, for GMCs associated with bright H II regions, when the FIR luminosity is computed from IRAS sky-flux images (Mooney \& Solomon 1988). Therefore we suggest a lower limit of $\sim 10\%$ for the contribution of embedded OB stars to the total FIR emission from GMCs undergoing massive star formation. Our estimate can be considered as a lower limit because: (a) some embedded OB stars are likely to illuminate dust regions so large in  angular size that they would not be identified as IRAS point sources and therefore not included in our sample and, (b) if identified, some of the source FIR flux may have been lost because of the background subtraction.

The ratio of FIR luminosity produced by embedded massive stars to virial mass for GMCs in the southern Galaxy, $\sim 0.4 \, L_{\odot}/M_{\odot}$, is similar to the FIR surface luminosity to H$_2$ surface density ratio found at the peak of the southern distribution of embedded OB stars, $\sim 0.41 \, L_{\odot}/M_{\odot}$. Therefore it is possible to suggest that most GMCs in that region of the Galaxy are presently forming massive stars, and it appears reasonable to use such value as a standard scale to evaluate the massive star formation efficiency per unit H$_2$ mass in the rest of the Galaxy relative to its maximum value. It is worth keeping in mind, though, that here we are averaging the H$_2$ surface density over large areas of the Galaxy, several kpc$^2$ in size. Therefore our maximum value for the FIR luminosity to H$_2$ mass ratio will be much smaller than that obtained when computed only for the neighborhood of a massive star forming region, with a typical FIR luminosity of $\sim 10^5\, L_{\odot} $ and a virial mass of $\sim 4 \, 10^3\, M_{\odot} $ (Plume et al. 1997)
 
The total FIR luminosity of $1.13\, 10^8\, L_{\odot}$ derived here for the embedded massive star layer, for $0.2 \leq R/R_{\circ} \leq 1.0$, represents $6.2\%$ of the FIR luminosity assigned to the H$_2$ layer by Bloemen et al. (1990) from IRAS data, $\sim1.83\,10^9\,L_{\odot}$, and $5.5\%$ of the FIR luminosity derived from ${\it COBE}$ DIRBE data ($\sim 2.05\,10^9\,L_{\odot}$) by Sodrosky et al. (1997). Because  in their respective analyses they consider all the FIR emission from the galactic plane, while here we compute only the FIR emission produced by dust fairly close to the heating stars, a strict lower limit of $\sim 6\%$ can be set for the contribution of embedded massive stars to the total FIR output from molecular gas within the solar circle.   

%Table6
\begin{table}
\
\caption{Separate axisymmetric models for the FIR luminosity
from UC H II regions: $R/R_{\circ}=0.2$ - 1.0}
\
\begin{tabular}{l@{\hspace{0.5cm}}l@{\hspace{0.5cm}}l}
	\hline
	\noalign{\smallskip}
	Parameter & North & South\\
	 & & \\ 
	\noalign{\smallskip}
	\hline
	\noalign{\smallskip}
      Total FIR luminosity (\pounds) & $9.43\,\,10^{7}\,L_{\odot}$ & $1.31\,
\,10^{8}\,L_{\odot}$\\
      Mean surf. lum. $<L>$ & $0.43\,L_{\odot}\,{\rm
pc}^{-2}$ & $0.60\,L_{\odot}\,{\rm pc}^{-2}$ \\
      Mean radius $<R>$ & $0.64\,R_{\circ}$ & $0.65\,R_{\circ}$ \\
      Mean half-width $<z_{\frac{1}{2}}>$ & 38 pc & 42 pc \\
      Mean displacement $<z_{\circ}>$ & $-6$ pc & $-3$ pc \\
     \noalign{\smallskip}
    \hline
\end{tabular}
\end{table}
 
%Table7
\begin{table}
\
\caption{Separate axisymmetric models for the H$_{2}$ distribution as traced 
by CO emission: $R/R_{\circ}=0.2$ - 1.0}
\
\begin{tabular}{l@{\hspace{0.5cm}}l@{\hspace{0.5cm}}l}
	\hline
	\noalign{\smallskip}
	Parameter & North & South\\
	 & & \\ 
	\noalign{\smallskip}
	\hline
	\noalign{\smallskip}
      Total H$_2$ mass$^{\rm a}$ ($M$) & $6.0\,\,10^{8}\,M_{\odot}$ & $6.2\,
\,10^{8}\,M_{\odot}$\\
      Mean surface density $<S>$ & $2.8\,M_{\odot}\,{\rm
pc}^{-2}$ & $2.9\,M_{\odot}\,{\rm pc}^{-2}$ \\
      Mean radius $<R>$ & $0.65\,R_{\circ}$ & $0.66\,R_{\circ}$ \\
      Mean half-width $<z_{\frac{1}{2}}>$ & 58 pc & 60 pc \\
      Mean displacement $<z_{\circ}>$ & $-12$ pc & $3$ pc \\
     \noalign{\smallskip}
    \hline
\end{tabular}
\begin{list}{}{}
\item[$^{\rm a}$] Using  $N({\rm H}_2)/W_{\rm CO} = 1.56\,10^{20}\,{\rm 
molecules}\,{\rm cm}^{-2}\,({\rm K}\,{\rm km}\,{\rm s}^{-1})^{-1}$ 
(Hunter et al. 1997) and $ R_{\circ} = 8.5 \,{\rm kpc}$  
\end{list}
\end{table}

\section 
{Conclusions}

Optical studies of massive stars are restricted to the solar
neighborhood due to extinction by intervening dust. The FIR spectral signature
of embedded OB stars provides a unique opportunity to extend the study of their
distribution to the whole galactic disk. We have analyzed here the first
complete galactic survey of dense molecular cloud cores associated with OB star
formation. Our main conclusions are: 

\begin{enumerate}

\item We have derived the mean radial distribution for 748 regions of massive star formation in the whole galactic disk. These regions produce a total FIR luminosity of $1.39\, 10^{8}\, L_{\odot}$ within the range $0.2 \leq R/R_{\circ} \leq 2.0$. We find 492 regions, representing $67\%$ of the sample, within the solar circle; they produce a FIR luminosity of $1.13\, 10^{8}\, L_{\odot}$, $81\%$ of the galactic total.   

\item Separate analyses of the 349 sources in the northern Galaxy, and of the 399 sources in the southern Galaxy, yield total FIR luminosities (extrapolated to the complete galactic disk) of $1.17\, 10^{8}\, L_{\odot}$ and of $1.60\, 10^{8}\, L_{\odot}$, respectively. The fraction of the total FIR luminosity produced within the solar circle ($\sim81\%$) is similar north and south.     
   
\item Regions of massive star formation in the galactic disk are distributed in a layer with its centroid $Z_{\circ}(R)$ following that of molecular gas for all galactocentric radii, both north and south. Within the solar circle the mean thickness of the OB star formation layer is $\sim 73$ pc (FWHM), a factor of 0.62 thinner than the H$_2$ layer ($\sim 118$ pc FWHM).
 
\item The FIR luminosity produced by massive stars, azimuthally averaged over the whole galactic disk within the solar circle, has a well defined maximum at $R = 0.55 \,R_{\circ}$, like the H$_2$ surface density, but with a gaussian FWHM of $0.28\,R_{\circ}$, a factor of 0.54 narrower than for the H$_2$ ($0.52 \,R_{\circ}$ FWHM).  

\item Toward the outer Galaxy, down from the maximum, the face-on FIR surface luminosity, azimuthally averaged over the whole galactic disk, decays with an exponential scale length of $0.21 \,R_{\circ}$, steeper than the H$_2$ surface density exponential decay which has a scale length of $0.34 \,R_{\circ}$.

\item Massive star formation per unit H$_2$ mass is higher than average in the region of the molecular annulus. While the mean FIR luminosity to H$_2$ mass ratio within the solar circle is $\sim 0.18\, L_{\odot}/M_{\odot}$, for $0.4 \leq R/ R_{\circ}  \leq 0.8$ the ratio grows to $\sim 0.23\, L_{\odot}/M_{\odot}$; and for $0.5 \leq R/R_{\circ} \leq 0.6 $ it reaches $\sim0.29\,L_{\odot}/M_{\odot}$  

\item Massive star formation per unit H$_2$ mass is maximum in the southern Galaxy at $0.5 \leq R/R_{\circ} \leq 0.6$. The average FIR luminosity to H$_2$ mass ratio there is $\sim 0.41\, L_{\odot}/M_{\odot}$, compared with $\sim 0.21\, L_{\odot}/M_{\odot}$ for its northern counterpart.

\end{enumerate}

There is a fair amount of further work that we hope will follow from the present results. The availability of a fairly complete and homogeneous sample of OB star forming regions in the Galaxy allows statistical studies of the general physical process of massive star formation, and in particular of the high-mass end of the IMF.  Bearing in mind that embedded massive stars should be the best tracers of spiral arm structure in our Galaxy, we foresee a concerted effort to resolve in a case-by-case basis their two-fold distance ambiguity within the solar circle.        

\begin{acknowledgements}  We are indebted to Dr. S. Digel for helping in
the selection of  sources; to F. Azagra for assistance with the data reduction;
to A. Luna for help with Fig. 5; and to Dr. E. Churchwell and Dr. T. Megeath for helpful discussions and suggestions. A thorough review by our referee greatly helped to improve the paper. The staff at the SEST and OSO telescopes kindly assisted us within  the course of the observations. The Swedish-ESO Submillimetre Telescope
is operated jointly by ESO and the Swedish National Facility for
Radioastronomy, Onsala Space Observatory, at Chalmers University of Technology.
The Onsala 20m telescope is operated by the Swedish National Facility for
Radioastronomy, Onsala Space Observatory, at Chalmers University of Technology.
L.B. acknowledges support from a Chilean {\it C\'atedra Presidencial en Ciencias}.   J.M., L.B., and S.C. acknowledge support from FONDECYT grant 8970017. 
  
\end{acknowledgements}

\end{document}